\documentstyle[prl,aps,multicol,amssymb,amsmath,epsf,psfrag]{revtex}

\def\unity{\mbox{\small 1} \!\! \mbox{1}}

\begin{document}

\title{The creation of large photon-number path entanglement\\
  conditioned on photodetection}
\author{Pieter Kok\cite{pieter}, Hwang Lee, and Jonathan P.\ Dowling}
\address{Quantum Computing Technologies Group, Section 367 \\
  Jet Propulsion Laboratory, California Institute of Technology \\
  Mail Stop 126-347, 4800 Oak Grove Drive, Pasadena, California 91109}

\maketitle

\begin{abstract}
 Large photon-number path entanglement is an important resource for enhanced
 precision measurements and quantum imaging. We present a general constructive
 protocol to create any large photon number path-entangled state based
 on the conditional detection of single photons. The influence of
 imperfect detectors is considered and an asymptotic scaling law is
 derived. 
\end{abstract}
\medskip


\begin{multicols}{2}

It has been known for some time now that quantum metrology techniques
allow for an
improvement in precision imaging and measurements by exploiting
entanglement. Examples of such improvements are in quantum lithography
\cite{boto00,kok01}, quantum gyroscopy \cite{dowling98},
entanglement-enhanced frequency metrology \cite{bollinger96}, and
clock synchronization \cite{jozsa00}. Experimental progress has been
made in the demonstration of lithography \cite{angelo01}, but
according to our present understanding, full-scale implementations
need sophisticated high-photon-number entangled states
\cite{kok01}. In particular, we need maximally entangled states of the form
$|N::0\rangle\equiv(|N,0\rangle+|0,N\rangle)/\sqrt{2}$, where $|N\rangle$ are
$N$-photon Fock states and $|0\rangle$ the vacuum. In general, we
use the following notation:
\begin{equation}
  |P::Q\rangle_{a,b}^{\varphi} \equiv \frac{1}{\sqrt{2}} \left(
   |P,Q\rangle_{a,b} + e^{i\varphi}|Q,P\rangle_{a,b}
   \right)\; ,
\end{equation}
where $a$ and $b$ denote the two subsystems (modes), and $\varphi$ is
a relative phase. There have been several proposals to
generate $|N::0\rangle$ states \cite{milburn89,gerry01}, but these typically
need materials with large $\chi^{(3)}$ nonlinearities of the order
of one. Currently known such non-linearities are very small; typically
they are of the order of $10^{-16}$ cm$^2$ s$^{-1}$ V$^{-2}$
\cite{boyd99}. 

In this paper we show how to create entangled states of large photon
number using only linear optics and photodetectors. In section \ref{parest},
we will give a brief overview of the theory of parameter estimation,
exploiting quantum entanglement to demonstrate the importance of the
$|N::0\rangle$ states. Then, in section \ref{path}, we present a
protocol to create $|N::0\rangle$ for any $N$. We show that it is
generalizable to arbitrary $N$ \cite{lee01}. In section \ref{det}, we
consider the case of imperfect detectors.

\section{Entanglement enhanced\\ parameter estimation}\label{parest}

In this section, we briefly describe the theory behind the various
entanglement enhanced imaging and measurement protocols. By using
results from parameter estimation theory, we may easily derive the
quantum noise limits for uncorrelated measurements, where every sample
is independent from every other, and for entanglement-enhanced
experiments, where events are corellated. Quantum lithography, though
technically not an estimation protocol, can also be described with
this theory. The main purpose of this section is to demonstrate the
importance of $|N::0\rangle$ states.

We start with the standard shot-noise limit. Consider an ensemble
of $N$ two-level systems in the state $|\varphi\rangle =
(|0\rangle+e^{i\varphi}|1\rangle)/\sqrt{2}$, where $|0\rangle$ and
$|1\rangle$ are arbitrary labels for the two levels. If $\hat{A} =
|0\rangle\langle 1| + |1\rangle\langle 0|$, then the expectation value
of $\hat{A}$ is given by  
\begin{equation}
  \langle\varphi|\hat{A}|\varphi\rangle=\cos\varphi\; .
\end{equation}
When we repeat this experiment $N$ times, we obtain
\begin{equation}
  _N\langle\varphi|\ldots\, _1\!\langle\varphi| \left(
  \,\overset{N}{\underset{k=1}{\mbox{\Large $\oplus$}}}\,
  \hat{A}^{(k)}\right)|\varphi\rangle_1 \ldots |\varphi\rangle_N = N
  \cos\varphi\; .
\end{equation} 
Furthermore, it follows from the definition of $\hat{A}$ that
$\hat{A}^2=\unity$ on the relevant subspace, and
the variance of $\hat{A}$ given $N$ samples is readily computed to be
$(\Delta A)^2 = N(1-\cos^2\varphi) = N \sin^2 \varphi$. According to
estimation theory \cite{helstrom76}, we have 
\begin{equation}\label{est}
  \Delta\varphi = \frac{\Delta A}{|d\langle
  \hat{A}\rangle/d\varphi|} = \frac{1}{\sqrt{N}}\; .
\end{equation}
This is the standard variance in the parameter $\varphi$ after $N$
trials. In other words, the uncertainty in the phase is inversely
proportional to the square root of the number of trials. This is the
shot-noise limit.

With the help of quantum entanglement we can improve this parameter
estimation by a factor of $\sqrt{N}$. We will now employ the path-entangled
input state $|N::0\rangle^{N\varphi}$, where $|N\rangle$ is a product 
collective state of the $N$ qubits. The relative phase
$e^{iN\varphi}$ can be obtained by a unitary evolution of one of the
modes of $|\varphi_N\rangle\equiv|N::0\rangle^{\varphi}$. When we
measure an observable $\hat{A}_N = |0,N\rangle\langle N,0| +
|N,0\rangle\langle 0,N|$ we have  
\begin{equation}\label{cosn}
  \langle\varphi_N |\hat{A}_N| \varphi_N\rangle = \cos N\varphi\; .  
\end{equation}
Again, $\hat{A}_N^2=\unity$ on the relevant subspace, and 
\begin{equation}
  (\Delta A_N)^2=1-\cos^2 N\varphi = \sin^2 N\varphi. 
\end{equation}
Using Eq.\ (\ref{est}) again, we obtain the so-called Heisenberg limit
to the minimal detectable phase: 
\begin{equation}\label{bol}
  \Delta\varphi_H = \frac{\Delta A_N}{|d\langle \hat{A}_N
  \rangle/d\varphi|}=\frac{1}{N}\; .
\end{equation}
Here, we see that the precision in $\varphi$ is increased by a
factor $\sqrt{N}$ over the standard noise limit, when we exploit quantum
entanglement. As shown in Bollinger {\em et al}.\ \cite{bollinger96},
Eq.\ (\ref{bol}) is the optimal accuracy permitted by the Heisenberg
uncertainty principle. In quantum lithography, one exploits the
$\cos(N\varphi)$ behaviour, exhibited by Eq.\ (\ref{cosn}), to print
closely spaced lines on a suitable substrate \cite{boto00}. Gyroscopy
and entanglement-enhanced frequency measurements
\cite{dowling98,bollinger96} exploit the $\sqrt{N}$ increased
precision. The physical interpretations of $A_N$ and the phase
$\varphi$ might differ in the different protocols.  

\section{large photon-number\\ path entanglement}\label{path}

In this section, we discuss the creation of large, photon-number path
entanglement using only linear optics and photodetectors. We first identify
the practical difficulties of conditioning on non-detection, and then,
instead, we introduce a generalizable scheme to create photon number
path entanglement based on actual detection. 

\subsection{Non-detection}

Previously, we have shown that it is possible to create up to
$|4,0\rangle + |0,4\rangle$ states with linear optics and projective
measurements \cite{lee01}. Subsequently, it was shown by
Fiur\'a\v{s}ek \cite{fiurasek01} and Zou {\em et al}.\ \cite{zou01}
that, in principle, one can create any two-mode, entangled,
photon-number eigenstate with linear optics and good Fock-state
sources. 

The difficulty with the Fiur\'a\v{s}ek-Zou protocols, however, is that
they are based on {\em non-detection}. There are two problems with
this approach: firstly, it means that the protocols are very sensitive
to detector losses; secondly, there is a whole family of reasons why a
detector will not register a photon (not necessarily connected to
detector efficiencies). For example, the lasers might have been
switched off, or the beams might be misaligned. In these cases there
will be no detector counts. In such situations the outgoing state is
not the required state but the vacuum.

More formally, let $|\Psi\rangle$ be the total state before any
detection and $|0\rangle_d\langle 0|$ the projection operator 
associated with a non-detection in mode $d$ (in this notation,
$|n\rangle_d\langle n|$ would be associated with the detection of $n$
photons). Furthermore, let $|\psi\rangle$ be the intended outgoing
state based on no photons in 
mode $d$. A perfect measurement of zero photons in mode $d$
corresponds to a projection $|0\rangle_d\langle 0|$, which yields a
state $|\psi\rangle\langle\psi|$. However, in practice the measurement
will not be a simple projection operator, but a positive
operator-valued measure (POVM) $\hat{E}_0$ \cite{kraus83}:
\begin{equation}
 \hat{E}_0 \equiv \sum_{n=0}^{\infty} c_{0,n} |n\rangle\langle n|\; ,
\end{equation}
where $c_{0,n}\geq 0$ and $\sum_k \hat{E}_k = \unity$. This will lead
to a different outgoing state:  
\begin{equation}
  \hat{\rho}_{\rm out} = {\rm Tr}[\hat{E}_0 |\Psi\rangle\langle\Psi|] =
  c_{0,0} |\psi\rangle\langle\psi| + (1-c_{0,0})\, \hat{\sigma}\; ,
\end{equation}
where $\hat{\sigma}$ is the density operator due to the noise. The
fidelity of the outgoing state is then given by $F = {\rm
  Tr}[\hat{\rho}_{\rm out}|\psi\rangle\langle\psi|] = c_{0,0}$. 

In general, there are many reasons why a detector might not record a
photon. Many of these can be tested (did I switch my equipment on?),
but never all of them. The crucial observation now is that all of the
{\em untested} possibilities are going to contribute to
$\hat{\sigma}$, and $c_{0,0}$ may become quite small
(see also Ref.\ \cite{kok00}).

This same argument can be applied to the detection of a single photon
(i.e., projecting mode $d$ onto $|1\rangle_d\langle 1|$). There will also
be a noise contribution in the form of a density operator analogous to
$\hat{\sigma}$. The difference is that there are many more reasons a detector
will not record the presence of a photon than there are for detecting
the photon. As a consequence, the fidelity of the outgoing state based
on detection will be much larger than the fidelity of the state based on
non-detection. 
 
When we have a low-fidelity output state, we need to apply
postselection. The output state therefore needs to be actually detected.
As long as we do not have suitable quantum non-demolition measurement
devices, the detection of the outgoing state generally precludes its
further use in the intended application. We therefore need a production
protocol that yields a high-fidelity output {\em state} (a notable
exception is quantum lithography, where states of different photon
number will not contribute to the imaging process \cite{boto00}). 

The question now is, what protocol allows us to create large
photon-number path entanglement conditioned upon photodetection? This
is the subject of the rest of the paper.

\subsection{Generating $|N::0\rangle$}

\begin{figure}[ht]
\epsfysize=2.0cm
\centerline{\epsfbox{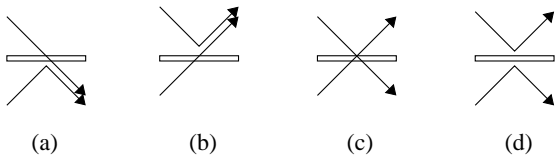}}
\vskip 12pt
\caption{Four possibilities exist when sending a $|1,1\rangle$ state through a
  beam splitter. The diagrams (c) and (d) lead to the same final
  state, but interfere destructively: (c) transmission-transmission
  $(i)(i)=-1$; (d) refection-reflection $(-1)(-1)=1$.}
\label{fig1}
\end{figure}

Let us first briefly recall the case of a two-fold coincidence at a
beam splitter \cite{lee01}. As shown in Fig.\ \ref{fig1}, when two
indistinguishable photons enter a 50:50 beam splitter in both input
modes, the phase relations will be such that the output modes will
always be in the state $|2::0\rangle$. This is the
operational mechanism of the Hong-Ou-Mandel (HOM) interferometer
\cite{hong87}. Labelling the input modes $a$ and $b$, and the output
modes $c$ and $d$, the beam splitter can be characterized by the
operator transformations 
\begin{eqnarray}
  \hat{a}^{\dagger} &\rightarrow& (-\hat{c}^{\dagger} + i
  \hat{d}^{\dagger})/\sqrt{2}\; , \cr 
  \hat{b}^{\dagger} &\rightarrow& (i\hat{c}^{\dagger} -
  \hat{d}^{\dagger})/\sqrt{2}\; ,  
\end{eqnarray}
and their Hermitean conjugates. It is now easily verified that
$\hat{a}^{\dagger}\hat{b}^{\dagger}$ transforms into
$(\hat{c}^{\dagger})^2 + (\hat{d}^{\dagger})^2$ up to an overall
phase. There are no cross terms, which is due to the reciprocal property 
of the symmetric beam splitter. However, this lack of cross terms will
not generalize beyond $N=2$, since there there are only a limited
number of free parameters available to suppress high-$N$ cross terms
\cite{dowling98a}. The critical property of the HOM interferometer, which we
will use in our protocol, is that two photons 
from different input modes of the beam splitter cannot trigger a
two-fold detection coincidence at the output modes. 

\begin{figure}[h]
  \begin{center}
  \begin{psfrags}
     \psfrag{x}{$c$}
     \psfrag{xx}{$c'$}
     \psfrag{y}{$d$}
     \psfrag{yy}{$d'$}
     \psfrag{a}{$a$}
     \psfrag{b}{$b$}
     \psfrag{c}{$a'$}
     \psfrag{d}{$b'$}
     \psfrag{e}{$\equiv$}
     \psfrag{f}{$\varphi$}
     \epsfxsize=8in
     \epsfbox[-10 10 890 150]{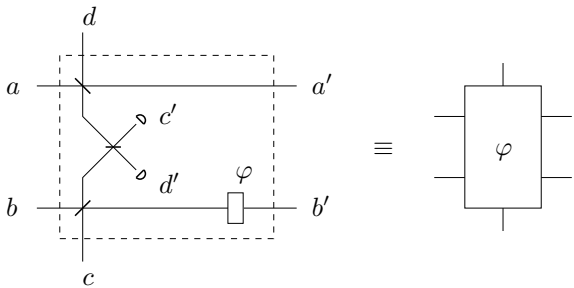}
  \end{psfrags}
  \end{center}
  \caption{The basic element of our large photon-number path
  entanglement generator. The beam splitters split off two photons,
  which are subsequently detected in a two-fold detector coincidence. 
  There is an extra phase freedom
  $\varphi$ in order to tune between several of such elements.}
  \label{fig2}
\end{figure}

In this section we first proceed with the general protocol for the creation of
$|N::0\rangle_{ab}$, where $N$ is even. The basic
element of our protocol is depicted in Fig.\ \ref{fig2}. Two beam
splitters split off photons from the main beams $a$ and $b$. The
reflected modes are then recombined in a 50:50 beam splitter, and the
process is post-selected on a two-fold detector coincidence in the
outgoing modes $c'$ and $d'$. It is assumed initially that our
detectors distinguish 
between one and more photons perfectly, but we consider the case of
imperfect detectors below. 

\begin{figure}[h]
  \begin{center}
  \begin{psfrags}
     \psfrag{f1}{$\varphi_1$}
     \psfrag{f2}{$\varphi_2$}
     \psfrag{fN}{$\varphi_{\frac{N}{2}}$}
     \psfrag{a}{$\!\!|N\rangle$}
     \psfrag{b}{$\!\!|N\rangle$}
     \psfrag{c}{$~|N::0\rangle$}
     \psfrag{v}{$\!\!|0\rangle$}
     \psfrag{dots}{\ldots}
     \epsfxsize=8in
     \epsfbox[-20 20 880 140]{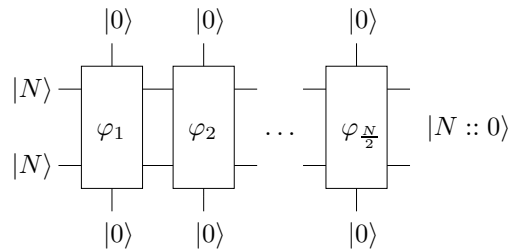}
  \end{psfrags}
  \end{center}
  \caption{Stacking the basic elements of figure 1 and setting
  the phase factors $\exp(i\varphi_k)$ to be the roots of unity, we
  create the state $|N,0\rangle+|0,N\rangle$ out of $|N,N\rangle$
  conditioned on an $N$-fold detection coincidence.}
  \label{fig3}
\end{figure}

Since a two-fold detector coincidence cannot be due to a single photon
in both input beams, this procedure thus takes two photons from either 
mode $a$ or $b$: $|N,N\rangle \rightarrow |N-2::N\rangle$. To complete
the element we apply a phase shift to mode $b$, the value of which we
will determine later. The protocol for making $|N::0\rangle$, with $N$
even now requires us to create the input state $|N,N\rangle$ and stack
$N/2$ of our basic elements. The output state, conditioned on an
overall $N$-fold detector coincidence with suitable phase shifts, is
then $|N::0\rangle$ (see Fig.\ \ref{fig3}). 

To prove this statement, consider the two-photon detection of the
basic element as $(\hat{a}^2 + e^{i\varphi}\hat{b}^2) |N,N\rangle$. We
have to repeat this procedure $N/2$ times, yielding
\begin{equation}\label{detection}
  \prod_{k=1}^{N/2} \left( \hat{a}^2 + e^{i\varphi_k}\hat{b}^2 \right)
  |N,N\rangle\; .
\end{equation}
In order to obtain the $N$-photon path-entangled state, the polynomial
in Eq.\ (\ref{detection}) should be $\hat{a}^N + \hat{b}^N$. This
means, from the fundamental theorem of algebra \cite{gauss}, that the
phase factors $\exp(i\varphi_k)$ are the roots of unity, that is,
\begin{equation}\label{phaseover2}
  \varphi_k = \frac{4\pi k}{N}\; .
\end{equation}

When $N$ is very large, the probability that only two photons are
reflected from the main beams becomes very small. This can be
compensated by the use of weighted beam splitters: To split off two
photons from an $N$-photon state most optimally, one should use a beam
splitter with a transmission coefficient $(N-1)/N$. The probability of
a successful state preparation event then scales asymptotically as
$\sqrt{8\pi N}(1/4e)^N$ (see appendix for proof).

So far, we have only considered the detection of an even number of
photons. However, for the general case, we also want to generate odd
$|N::0\rangle$ states. The even case was straightforward,
since it involved only two-photon detections, which are naturally
implemented as the detection of the two outgoing modes of a
beamsplitter. The odd case, however, requires single photon
detectors. If we allow for non-detection, this is also a
straightforward task ---we just condition on a single detection count in
the two outgoing modes of the beam splitter. But non-detection is exactly what
we want to avoid. In the next section, we investigate 
single-photon conditioning in the presence of polarization.

\subsection{Odd $N$ and polarization degrees of freedom}

The protocol presented in the previous section generates only even-$N$
path-entangled states $|N::0\rangle$. Furthermore, the
photons are assumed to have the same polarization. In this section, we
extend this scheme to odd $N$ by using the extra degree of freedom of
polarization. 

\begin{figure}[h]
  \begin{center}
  \begin{psfrags}
     \psfrag{x}{$c$}
     \psfrag{xx}{$c'$}
     \psfrag{y}{$d$}
     \psfrag{yy}{$d'$}
     \psfrag{a}{$a$}
     \psfrag{b}{$b$}
     \psfrag{c}{$a'$}
     \psfrag{d}{$b'$}
     \psfrag{f}{$\varphi$}
     \epsfxsize=8in
     \epsfbox[-100 0 800 175]{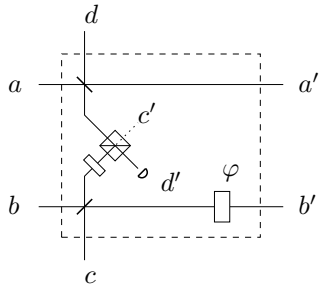}
  \end{psfrags}
  \end{center}
  \caption{This element is used to subtract a single photon from the
     two-mode entangled state. One photon, originating from either
     mode $a$ or $b$, will either be transmitted in the polarization
     beam splitter to the detector (mode $a$), or it will undergo a
     $\pi/2$ polarization rotation and will reflected to the
     detector (mode $b$). The second outgoing mode is
     empty. Rather than conditioning on a non-detection of the
     empty mode, we couple it to the environment.} 
  \label{fig4}
\end{figure}

The basic element for subtracting a photon from the main modes is
shown in Fig.\ (\ref{fig4}). Just as with the even case, above, two beam
splitters split off a portion of the main beams $a$ and $b$. However, now
 they are recombined in a polarization beam splitter (PBS). The
setup is chosen such that a photon originating from mode $a$ will be
transmitted in the PBS. Since the polarization of modes $a$ and $b$ are
the same, a photon from mode $b$ incident on the PBS would also be
transmitted. However, we really want this photon to be reflected, so
that it too ends up in the detector. In this way we erase the which-path 
information. 

One way to achieve this goal is to apply a $\pi/2$ polarization
rotation to this mode. This will force the photon towards the 
detector. The secondary
outgoing mode will now be empty, and as a consequence we will ignore
it completely. That is, we do not need to condition this scheme on
non-detection. When we stack these elements $N$ times and use the
input state $|N,N\rangle$, we create a general $|N::0\rangle$
state. The phase factors $\exp(i\varphi_k)$ are the $N$
roots of unity: 
\begin{equation}
  \varphi_k = \frac{2\pi k}{N}\; .
\end{equation}
Note that, as in Eq.\ (\ref{phaseover2}), the phases span the
$N/2$ roots of unity. 

Let us elaborate a bit more on this distinction between non-detection
and losing modes to the environment. In any experiment we trace out
the (unwanted) coupling to the environment, for the simple reason that
we do not have control over all the interactions between our
experimental setup and the rest of the universe. When this coupling is
made small (i.e., the setup is isolated), this is a very good
approximation. In the 
case of the secondary output mode $c'$ of the PBS, the coupling to the
environment is the loss of any photons in that mode. However,
ideally there should not be any photons in that mode ---this therefore
constitutes a weak coupling.

The case of non-detection presupposes a sizeable portion of scattered
photons in the outgoing beam, and aims to condition on the absence of
these. We cannot trace over this mode, because the coupling to the
environment is not weak ---there are actually photons in that mode. This
means that instead of tracing over the secondary mode, one needs to
project it onto the vacuum $|0\rangle\langle 0|$, which is the source
of the non-detection difficulties.

\subsection{Nested protocols}

The protocols presented so far are linear, in the sense that we
constructed a basic element as a two-mode gate that was repeated a
number of times. This means that the number of detected photons
increases linearly with the number of elements, and the efficiency
(which is a product of the success rate of the different components)
therefore scales exponentially poorly.

For practical purposes it is important to find a scheme which scales
logarithmically in the number of detectors, so that we only have
polynomial efficiency deterioration. One protocol that looked
promising exploited the unused input ports. We found that feeding both
modes $a,b$ {\em and} modes $c,d$ in the basic element from Fig.\
\ref{fig2} with states $|N::0\rangle$ yields the state
\begin{equation}
  |\psi_{\rm out}\rangle = |2N-2::0\rangle\; ,
\end{equation}
based on a two-fold detection coincidence. However, due to the fact
that {\em two} $|N::0\rangle$ input states are required,
the overall scaling was still exponentially poor.

These scaling considerations are important for practical implementations of
entanglement-enhanced precision measurements, because an increase in
the required resources (photons) might outweigh the benefit of gaining
a $\sqrt{N}$ precision improvement. Since the scaling of the resources depends
critically on the details of the protocol employed, it is not clear from
these general considerations what the overall behaviour of a given
network will be.

\section{imperfect detectors}\label{det}

There are several sources of errors for a detector. It might fail to
signal a photon was present, in which case we speak of a
deteriorated efficiency. Alternatively, it might signal the detection
of a photon, even though no photon was actually present. This is
called a dark count. Since we only consider schemes that operate in
short time windows, these dark counts can be neglected. Finally,
the detector might not be able to distinguish between one or more
photons. Such a detector does not have single-photon resolution
\cite{kok01a}. 

We can see immediately that imperfect detection efficiency is going to
affect the scaling law. In particular, the asymptotic scaling will
behave as
\begin{equation}
  \pi_N = \sqrt{8\pi N}\left(\frac{\eta}{4 e}\right)^N 
\end{equation}
where $\pi_N$ is the (asymptotic) probability of creating the state
based on $N$ detected photons and $\eta$ is the detector
efficiency. That is, the protocol scales exponentially poorly with the
detector efficiency, as expected. Here we have taken identical
detectors throughout the scheme. 

When we use detectors with a single-photon resolution but limited
efficiency, two photons can easily be mistaken for a single
photon. That is, one of them might not be detected. When the
occurrence of a two-photon state is very unlikely, this is not so much
of a problem, but when it is likely, the output state will be
significantly degraded. Unfortunately, in our protocol the beam
splitter strips off two photons on average, which means that it is
quite likely that more than two photons end up in the detector.
This way, our scheme has become a protocol conditioned on the
non-detection of two-photon states, which is exactly what we
wanted to avoid.

However, there is a way to mend this drawback: when we increase the
transmittivity of the beam splitter, the probability of having more
than two photons in the detectors will decrease. Therefore, at the
cost of a lower production rate (i.e., low efficiency), we can
maintain high-quality $|N::0\rangle$ states (high fidelity).
This adjustment is not possible in non-detection schemes.

\section{conclusions}

In this paper we have demonstrated a general, detection-based protocol
to create $|N::0\rangle$ states for use in
entanglement-enhanced parameter estimation. Existing protocols are
less practical because they either require $\chi^{(3)}$ nonlinearities
near unity \cite{milburn89,gerry01}, or they condition on non-detection
\cite{fiurasek01,zou01}. Currently, $\chi^{(3)}$ nonlinearities are
very small \cite{boyd99}, and we argued that non-detection schemes are
problematic in their experimental implementation. 

We have shown that one can indeed create arbitrary $|N,0\rangle +
|0,N\rangle$ states using only linear optics and conditioned on
single-photon detections. For the case of odd $N$ we needed to invoke
the extra freedom of polarization. The protocol presented here is the
generalization of our previous work \cite{lee01}, which was succesful
in creating path-entangled states up to $|4,0\rangle+|0,4\rangle$.

\section*{acknowledgements}

This work was carried out at the Jet Propulsion Laboratory, California
Institute of Technology, under a contract with the National
Aeronautics and Space Administration. In addition, P.K.\ and H.L.\
acknowledge the United States National Research Council. The authors
would also like to thank N.\ Cerf, J.D.\ Franson, G.\ Hockney, P.G.\
Kwiat, G.J.\ Milburn, Y.H.\ Shih, D.J.\
Wineland, and U.H.\ Yurtsever for useful discussions. 
Support was received from the Office of Naval Research, Advanced Research
and Development Activity, National Security Agency and the Defense
Advanced Research Projects Agency. 

\appendix

\section*{asymptotic scaling law}

We will now prove the asymptotic scaling law for even
$|N,0\rangle + |0,N\rangle$ states. Let the transmission and
reflection coefficient of the beam splitter be given by $t$ and $r$,
with $t+r=1$. The probability of reflecting $k$ out of $N$ photons is
then given by $p_k(N)=\binom{N}{k} t^{N-k} r^k$, from which it
immediately follows that $\sum_k p_k(N)=1$. The event where two
photons are reflected from one beam splitter and none from the other
then occurs with probability $2 p_0(N) p_2(N)$, where the factor of 2
takes into account the fact that we do not know from which mode the
two photons originate. Furthermore, we are post-selecting on two-fold
coincidences, which means that we have an extra factor of
$\frac{1}{2}$ that incorporates the reduced probability that
the two photons branch off at the beam splitter. By maximizing the
expression $t^{2N-2}(1-t)^2$ we found that the optimal transmission
coefficient is given by $t=(N-1)/N$. The total probability of finding
a two-fold detector coincidence is then given by  
\begin{equation}\nonumber
 p_{\text{two-fold}} = 2 \left(\frac{1}{4}\right)^N \frac{N!}{N^N}\; .
\end{equation}
Using Sterling's formula $N! \approx \sqrt{2\pi N} e^{-N} N^N$, we
find that for large $N$ the protocol scales as $\sqrt{8\pi N} (1/4e)^N$.

\end{multicols}
\end{document}